# Realization of large magnetocaloric effect in the Kagome antiferromagnet $Gd_3BWO_9$ for Sub-Kelvin cryogenic refrigeration

Fangyuan Song[1*], Xinyang Liu[2*], Chao Dong[1], Jin Zhou[1], Xinlong Shi[1], Yuyan Han[3], Langsheng Ling[3], Huifen Ren[2], Songliu Yuan[1], Shun Wang[1,4], Junsen Xiang[2≡], Peijie Sun[2≡], Zhaoming Tian[1 #]

[1] Wuhan National High Magnetic Field Center and School of Physics, Huazhong University of Science and Technology, Wuhan, 430074, China

[2] Beijing National Laboratory for Condensed Matter Physics, Institute of Physics, Chinese Academy of Sciences, Beijing, 100190, China

[3] Anhui Key Laboratory of Low-energy Quantum Materials and Devices, High Magnetic Field Laboratory, Chinese Academy of Sciences, Hefei, Anhui 230031, China

[4] School of Fundamental Physical quantum Physics, Huazhong University of Science and Technology, Wuhan, 430074, China

Rare-earth (RE) based frustrated magnets have attracted great attention as excellent candidates for magnetic refrigeration at sub-Kelvin temperatures, while the experimental identification on systems exhibiting both large volumetric cooling capacity and reduced working temperatures far below 1 K remain to be a challenge. Here, through the ultra-low temperature magnetism and thermodynamic characterizations, we unveil the large magnetocaloric effect (MCE) realized at sub-Kelvin temperatures in the frustrated Kagome antiferromagnet $Gd_3BWO_9$ with $T_N$~1.0 K. The isothermal magnetization curves indicate the existence of field ($B$) induced anisotropic magnetic phase diagrams, where four distinct magnetic phases for $B$ // $c$-axis and five magnetic phases for $B$ // ab-plane are identified at $T$< $T_N$. The analysis of magnetic entropy $S(B, T)$ data and direct adiabatic demagnetization tests reveal a remarkable cooling performance at sub-Kelvin temperatures featured by a large volumetric entropy density 502.2 mJ/K/cm$^3$ and a low attainable minimal temperature $T_{min}$~168 mK from the initial cooling condition of 2 K and 6 T, surpassing most of Gd-based refrigerants previously documented in temperature ranges of 0.25-4 K. The realized $T_{min}$~168 mK far below $T_N$ ~ 1.0 K in $Gd_3BWO_9$ is related to the combined effects of magnetic frustration and criticality-enhanced MCE, which together leave a substantial magnetic entropy at reduced temperatures by enhancing spin fluctuations.

**Keywords:** magnetocaloric effect, frustrated magnets, quantum criticality, phase diagram

---

* These authors contributed equally to this work.

≡ Corresponding author: xiangjs@iphy.ac.cn

≡ Corresponding author: pjsun@iphy.ac.cn

# Corresponding author: tianzhaoming@hust.edu.cn

## 1. Introduction

Experimental realization of sub-Kelvin cooling is imperative for both fundamental research and modern technology applications. Adiabatic demagnetization refrigeration (ADR) based on the magnetocaloric effect (MCE) of magnetic refrigerants [1, 2], provides a valuable technique for achieving ultra-low temperatures. In comparison with the prevailing cryogenic refrigeration based on the liquid helium ($^3$He-$^4$He) dilution refrigeration, ADR cooling technique has the advantages such as cryogen-free operation, environmental sustainability, compact design and ease of operation, etc [3-5]. Thus, this technique has attracted growing attention for its potential of widespread applications in solid-state based quantum devices/sensors and astronomical cryogenic systems under microgravity conditions [6, 7]. Currently, commercial magnetic refrigeration down to the sub-Kelvin temperatures uses the dilute paramagnetic salts with very low ordering temperature ($T_o$) such as $Ce_2Mg_3(NO_3)_{12} \cdot 24H_2O$ (CMN) [8] and $Mn(NH_4)_2(SO_4)_2 \cdot 6H_2O$ (MAS) [9]. Here, the attainable minimum temperature ($T_{min}$) is limited by the $T_o$ of magnetic refrigerants. However, the low volumetric magnetic ion density ($N_{mag}$) diluted by the nonmagnetic ions restrict their cooling capacity and specific power. This limitation has driven the identification of novel type of efficient ADR materials working at sub-Kelvin temperatures, which need to combine a large $N_{mag}$ and a low $T_o$. On the other side, larger $N_{mag}$ of ADR materials usually lead to a higher $T_o$ and correspondingly smaller residual magnetic entropy at low temperatures since the entropy release happens at a narrow temperature window near $T_o$. Recently, this dilemma has been partially solved by employing the geometrically frustrated magnets (GFMs) as the ADR coolants [10-15], where $T_o$ can be suppressed by geometrical frustration effect and strong spin fluctuations, leaving a substantial ground state entropy at reduced temperatures [11,15-19].

In comparison with the 3d transition-metal based GFMs, rare-earth (RE) based compounds are more competitive ADR coolants in sub-Kelvin temperatures due to the intrinsic weaker exchange interactions among the 4$f$ local moments, enabling them to easily reach lower working temperatures in the same $N_{mag}$. Among them, Gd-based GFMs are especially attractive for improving the cooling capacity due to the large magnetic entropy $S_M$=$R$ln(2S+1)=$R$ln8 in contrast to the $S_M$=Rln2 in the other RE-based compounds that exhibit effective spin-1/2 moments, since only the entropy of crystal-electric-field (CEF) ground state of RE ions can be effectively used in this temperature window [16]. Indeed, the recent cooling performance tests have revealed that Gd-based coolants can possess a larger magnetic entropy change ($\Delta S_M$) and a longer cooling duration time at temperatures around $T_{min}$ than the isostructural Yb-based compounds[11,18-21], despite the much lower $T_o$ for the latter. As typical examples, the frustrated hyperkagome lattice $Gd_3Ga_5O_{12}$ (GGG) [4,22,23] and triangular lattice $KBaGd(BO_3)_2$ [18,19] compounds have been revealed to be excellent ADR coolants, which can achieve $T_{min}$ ~ 322 mK and $T_{min}$~ 122 mK upon demagnetization from the initial conditions of 2 K and 7 T. Stimulated by the effectiveness of spin frustration on enhancing sub-Kelvin refrigeration, other Gd-based frustrated magnets including pyrochlore-lattice $Gd_2Ti_2O_7$ [24], face-centered cube (FCC)-lattice $Ba_2GdSbO_6$ [15] and frustrated spin chain $Gd_{9.33}[SiO_4]_6O_2$ [25] have also been identified. To further promote the cooling performance of Gd-based magnets, it is essential to seek for new types of frustrated lattice systems with high $N_{mag}$ and the suitable working temperature window with excellent cooling performance. Meanwhile, new experimental approaches by utilizing novel magnetic states and enhanced MCE near quantum criticality are essential for advancing magnetic cooling.

Kagome-lattice antiferromagnets as typical GFMs with high degree of geometric frustration, can be used as ADR coolants. As one candidate, the recently discovered RE-based kagome compounds $RE_3BWO_9$ (RE=Pr-Ho) have been intensely studied for hosting exotic magnetic states [26]. In these series, magnetic $RE^{3+}$ ions occupy on the kagome lattices and are stacked in the "ABAB"-type fashion along the *c*-axis [see in Fig.1(a) ]. In $Gd_3BWO_9$, the small intralayer and interlayer's Gd-Gd distances (3.872-4.799 Å) indicate a magnetically dense structure enabling a large cooling capacity. Here, we report the low-temperature anisotropic magnetism, specific heat and ADR cooling performance of $Gd_3BWO_9$ antiferromagnet ordered at $T_N$~ 1 K. Based on the systematical magnetization measurements, the field-temperature magnetic phase diagrams are established. More importantly, both the analysis of magnetic entropy data and direct ADR test reveal $Gd_3BWO_9$ to be an excellent MCE material, featured by the combination of a large volumetric magnetic entropy, an attainable quite low $T_{min}$~168 mK and a long holding time after the ADR process from the initial condition of 2 K and 6 T.

## 2. Experimental section

Single crystals of $Gd_3BWO_9$ were grown using a flux method as described in previous reports [27, 28], the transparent single crystals with typical dimensions of 5mm×2mm×1mm can be obtained [see Fig. 1(b)]. The crystal structure was determined by Bruker single-crystal X-ray diffraction (SXRD) using graphite-monochromated Mo K$α$ radiation ($λ$ = 0.71073 Å), the refined crystallographic data are listed in Tables S1-S3 in the Supplemental Material [29]. Using the crushed single crystals, powder X-ray diffraction (XRD) spectra were collected in a PAN Analytical X'Pert Pro MPD diffractometer with Cu K$α$ radiation ($λ$ = 1.5418 Å), the refined XRD spectra is shown in Fig. S1. From that, we extract the information on lattice parameters and crystal structure.

Temperature dependence of magnetic susceptibility $χ(T)$ and isothermal field ($B$) dependent magnetization $M(B)$ measurements between 1.8 K and 300 K were conducted using the Quantum Design superconducting quantum interference device magnetometer. At temperatures down to 0.4 K, magnetic characterizations were carried out by the MPMS3 equipped with a $^3$He option. Specific heat data were collected using the physical properties measurement system (PPMS, Quantum Design) at different magnetic fields. The ultra-low temperature specific heat with temperature down to 80 mK was measured in PPMS using the heat capacity option equipped with a dilution refrigeration system.

To characterize magnetic cooling performance in sub-Kelvin temperatures, the quasi-isentropic temperature change with varying magnetic field and holding time at zero field of the samples were performed on the PPMS based setup by using a homemade sample stage, as described in Refs. [13,30]. Both $Gd_3BWO_9$ single crystals with a mass of 0.7 g and polycrystals with a mass of 3 g samples were employed for the measurements, and the silver powders with a mass of 1.5 g were mixed with the polycrystalline samples to improve the thermal conductivity. During the quasi-adiabatic demagnetization (ADR) measurements, the initial cooling conditions were set to $B_i$= 2-7 T and $T_i$ = 2 K. Additionally, at temperatures above 1.8 K, the adiabatic temperature changes in response to field have been measured using pulsed magnetic field under quasi-adiabatic conditions [31,32], which were carried out at the Wuhan National High Magnetic Field Center. Pulsed high fields up to 32 T were generated by a long-pulse magnet with duration of 80 ms, and the exact temperature was detected by a thin film $RuO_2$ thermometer on the surface of the sample. Due to the small and monotonic magnetoresistance response in our measured field ranges of $RuO_2$ film, $RuO_2$ thermometer can be calibrated to ensure the accuracy of measurement [31]. Additionally, the $Au_{16}Ge_{84}$ thermometer is used for the measurement and which gives the same magnetocaloric data in pulsed high fields. At various based temperature ($T_o$), the temperature ($T_s$) of thermometer attached on the $Gd_3BWO_9$ samples versus magnetic field were obtained for field applied along the *c*-axis.

## 3. Results and discussion

### 3.1 Magnetic ground state

$Gd_3BWO_9$ crystallizes in the hexagonal structure with space group $P6_3$. Based on the structural refinements of XRD data, the resultant lattice parameters are *a* = *b* = 8.557(3) Å and *c* = 5.4082(18) Å, respectively. The structure analysis confirms a single phase of the synthesized single crystals. No distinct antisite occupations between magnetic $Gd^{3+}$ and non-magnetic $B^{3+}/W^{6+}$ cations are detected, which can be due to the large differences of ionic radii and distinct local oxygen coordination environments between $Gd^{3+}$ and $B^{3+}/W^{6+}$ cations. It is found that the magnetic $Gd^{3+}$ ions form $GdO_8$ dodecahedrons with low local symmetry (point symmetry $C_1$). The magnetic $Gd^{3+}$ ions form a distorted Kagomé lattice in the *ab*-plane that are stacked in the "AB" type fashion along the *c*-axis [see Fig. 1(a)]. Within the Kagome plane, the two $Gd^{3+}$ regular triangles have different Gd-Gd bond lengths of $r_1^{intra}$=4.228 Å and $r_2^{intra}$=4.799 Å slightly larger than the interlayer spacing separation of $r^{inter}$ =3.872 Å, in accordance with the results of polycrystals [26]. Herein, the small nearest neighboring Gd-Gd distances enable a large magnetic ion density $N_{mag}$=17.49×$10^{21}$ cm$^{-3}$, ~40% larger than the hyperkagome lattice $Gd_3Ga_5O_{12}$ system [22].

Figure. 1(c) shows the isothermal magnetization $M(B)$ curves at 1.8 K for *B* along the *c*-axis and *ab*-plane, respectively. For field oriented within the *ab*-plane, the *b**-axis is selected as schematically shown in Fig. 1(c). A weak magnetic anisotropy is observed between the two field directions, in accordance with the magnetic susceptibility $\chi$(T) data, which nearly overlap at temperatures above 2 K [ see Fig. 1(d)]. In temperature range of 5 K ≤ *T* ≤ 15 K, the Curie-Weiss (CW) fitting on the inverse magnetic susceptibility $1/\chi(T)$ yields the effective magnetic moments $\mu_{\text{eff,c}}$= 8.31 $\mu_B$/Gd ($B$ // $c$-axis) and $\mu_{\text{eff,b*}}$ = 8.28 $\mu_B$/Gd ($B$ // $ab$-plane), which are slightly larger than the free ion moments $\mu_{eff}^{\text{free}} = 7.94\ \mu_B$ of $Gd^{3+}$ ions ($4f^7$, $S=7/2, L=0$). The negative CW temperatures $\theta_{\text{CW,c}}$ = -0.87 K ($B$ // $c$-axis) and $\theta_{\text{CW,b*}}$ = -1.13 K ($B$ // $ab$-plane) reveal the dominant antiferromagnetic (AFM) interactions between the local moments of $Gd^{3+}$ ions.

To unveil the magnetic ground state, zero-field specific heat $C_p(T)$ down to 80 mK is shown in Fig. 2(a). A sharp peak at $T_N$~1.0 K suggests the onset of long-range antiferromagnetic (AFM) order of $Gd_3BWO_9$. After subtracting the lattice contribution $C_{\text{Latt}}$ using the specific heat of nonmagnetic analog $Eu_3BWO_9$ as background, the resultant magnetic specific heat $C_M(T)$ is presented in Fig. 2(b). The magnetic entropy $S_M(T)$ can be obtained by integrating $C_m/T$ versus $T$, which approaches saturation of $R\ln8$ at 5 K as expected for $Gd^{3+}$ ions with $S$ = 7/2. Also, we can notice that large magnetic entropy ~$0.6R\ln8$ is released well below $T_N$~ 1 K, rendering $Gd_3BWO_9$ suitable for sub-Kelvin magnetic refrigeration. Below $T_N$, $C_M(T)$ curves display a crossover behavior at temperature $T^*$~0.37 K, above $T^*$ the specific heat follow a power law $C_M(T) \propto T^{1.5}$ (0.37 K < $T$ <1.0 K) and below that it changes to $C_M(T) \propto T^2$ ($T$ < 0.37 K) behavior [see the inset of Fig. 2(a)]. This is different from the conventional three-dimensional antiferromagnets following the $C_M(T) \sim T^3$ dependence. Here, $C_M(T)$ data drops out slower than the $T^3$ dependence, thus lead to larger amount of entropy released at temperatures below $T_N$. Besides, the $C_M(T) \propto T^2$ dependence at $T < T^*$ suggests the possible novel low-energy excitation with persistent spin dynamics as the report in other frustrated Gd-based systems including the hyper-Kagome lattice $Gd_3Te_2Li_3O_{12}$ and pyrochlore lattice $Gd_2Pb_2O_7$ [33-35]. The observed magnetic crossover indicates a possible change of spin dynamics above and below $T^*$ ~ 0.37 K.

**3.2 Magnetic phase diagrams**

The isothermal magnetization $M(B)$ and its field derivative magnetization $dM/dB$ curves at 0.4 K of $Gd_3BWO_9$ are presented in Figs. 3(a) and 3(b) for field along the $c$-axis and $b^*$-axis, respectively. Along both directions, magnetization approaches its saturation $M_S$= 7.0 $\mu_B/Gd^{3+}$ at high fields. Above ~ 2 T, the nearly overlapped $M(B)$ curves indicate the isotropic field responses [see Fig. S2]. This is expected for the $Gd^{3+}$ ions with the $4f^7$ half-filled configuration (S=7/2, $L$ = 0). At low field regions ($B$ < 1 T), $M(B)$ curves exhibit anisotropic field responses. For $B$ // $ab$-plane, three magnetic anomalies are observed at critical fields $B_{\text{ab,1}}$~0.045 T, $B_{\text{ab,2}}$~0.27 T and $B_{\text{ab,3}}$~0.55 T, where $B_{\text{ab,1}}$, $B_{\text{ab,2}}$ and $B_{\text{ab,3}}$ are determined by the peak of the $dM/dB$ curves in Fig. 3(b). At $B_{\text{ab,2}}$ and $B_{\text{ab,3}}$, magnetization reaches ~$1/3M_s$ and ~$2/3M_s$, respectively. By comparison, for $B$ // $c$-axis, two magnetic anomalies appear at fields $B_{c1}$~0.35 T and $B_{c2}$~0.82 T, and magnetization at $B_{c2}$ is larger than $2/3M_s$. As field increases above the saturated field ($B_{c,s}$), magnetization approaches its saturation.

To more clearly illustrate the magnetic behaviors, the field derivative $dM/dB$ curves at selected temperatures along both directions are shown in Figs. 3(c) and 3(d), respectively. At temperatures near or above $T_N$ = 1 K, a single broad hump is observed in the d$M$/d$B$ curves with the maximum located at a critical field ($B_{cr}$) characterizing a field-induced magnetic crossover, since short-range spin correlation has been developed in this temperature region from the analysis of $M(B)$ results ( see Fig. S2). Below $T_N$, several sharp anomalies appear in the d$M$/d$B$ curves, signifying the field-induced successive magnetic transitions. Based on the d$M$/d$B$ data, the contour plots of field-temperature ($B$-$T$) are constructed as shown in Fig. 3(e) and 3(f), respectively. The intensity of the contour plot corresponds to the values of d$M$/d$B$ data, which well depicts all the magnetic phase boundaries. At zero field, a long-range AFM ground state is established below $T_N$. Upon applying field along the $c$-axis, two metamagnetic transitions at critical fields ($B_{c1}$, $B_{c2}$) are observed at temperatures below 0.88 K, between which an intermediate AFM phase (IAFM1) can be identified. Alongside the magnetic crossover line determined by $B_{cr}$, the above three critical lines converge at a multiple critical point located at $B_{cp}$~0.41±0.02T and $T_{cp}$~0.89 ±0.01 K in the phase diagram. Considering that $B$ is oriented out of the kagome plane, the IAFM1 phase likely corresponds to a non-coplanar spin configuration, similar to that observed in its isostructural Nd$_3$BWO$_9$ [36, 37] and triangular lattice Na$_2$BaCo(PO$_4$)$_2$[13]. As $B$ > $B_{c2}$, it enters into another intermediate magnetic state (IAFM2), Gd$^{3+}$ moments gradually align along the $c$-axis. Above $B_{c,s}$, it becomes fully polarized state (FP). While, for $B$ // $ab$-plane, five distinct magnetically ordered phases are identified at T < $T_N$ [see Fig. 3(f)]. Moreover, since the observed fractional magnetizations 1/3$M_s$ and 2/3$M_s$ are usually ascribed to the formation of "up-up-down" and "up-up-up-up-up-down" spin configurations in the triangular lattice, the inferred low-field AFM (LAFM) ordered state is expected to evolve into a coplanar canted "Y-shaped" ($B_{ab,2}$ < $B$ < $B_{ab,3}$) and canted 'V-shaped' spin structure ($B_{ab,3}$< $B$ < $B_{ab,s}$) under magnetic field in the $ab$-plane. For $B$ > $B_{ab,s}$, it enters into the FP state. For the observed fractional magnetizations, since the Gd$^{3+}$ ions have large $S$=7/2 effective moments with weak quantum spin fluctuations, large thermal fluctuations possibly play an important role on their appearance, similar to the report in triangular-lattice RbFe(MoO$_4$)$_2$ ($S$=5/2) [38] and Kagome lattice Cs$_2$KCr$_3$F$_{12}$ (S=3/2) magnets [39].

### 3.3 Specific heat and magnetocaloric effect

The low-temperature $C_M$(T) curves of Gd$_3$BWO$_9$ at selected fields for $B$ // $c$ are presented in Fig. 4(a). Upon increasing $B$, the peak in the $C_M$(T) curves shifts to low temperatures and is gradually smeared out at $B$~1.5 T. At low magnetic fields ($B$ ≤ 1.2 T), a crossover behavior is observed in $C_M$(T) curves at temperature $T^*$~0.37 K. Below $T^*$, the $C_M$/T data show a linear temperature dependence with a slight decrease in slope as field increases. Above 1.5 T, no sharp peak but only a broad Schottky-like anomaly emerges, with its position gradually shifting to higher temperatures [see Fig. 4(b)]. A similar trend is observed for $B$ // $ab$-plane, the C$_M$(T) curves at high fields of 3 T and 7 T are nearly overlapped with the ones for $B$ // $c$, indicating the nearly isotropic field response in polarized magnetic states ($B$ ≥2 T).

Figure. 4(c) displays the resultant magnetic entropy $S_M$($T$,$B$) at different fields, obtained by integrating $C_M(T)/T$ versus temperature. The $S_M(T,B)$ gives the evolution of magnetic entropy of $Gd_3BWO_9$ at different $T$ and $B$, which can be used to estimate the MCE and minimum temperature attainable in the adiabatic demagnetization refrigeration (ADR) process. At low fields ($B ≤ 1$ T), the $S_M(T, B)$ curves approach saturation as $T$ increases up to 20 K with $S_M=R\ln 8$, indicative of full release of magnetic entropy of $Gd^{3+}$ ions. From the $S_M(T, B)$ data, the magnetocaloric properties can be evaluated as illustrated by the grey arrows, which include an isothermal entropy change ($-\Delta S_M$) in a given field change $\Delta B = B_i-B_0$ and adiabatic temperature change $\Delta T_{ad} = T_i-T_f$ in the isentropic process. In Fig.4(c), $B_0 = 0$ T and $B_i = 7$ T are selected for demonstration, with $T_i$ and $T_f$ marking the initial and final temperature in the ADR process from the initial condition ($T = T_i$, $B = B_i$). Here, the $-\Delta S_M$ characterizes the cooling capacity of refrigerants as the driving force in the ADR process, and $T_f$ refers to the attainable temperature as field is driven to zero. Setting $T_i = 2$ K and $\Delta B = 7$ T [see Fig. 4(d) and Fig. 4(f)], $Gd_3BWO_9$ shows a nearly isotropic magnetic cooling performance for $B$ // c and $B$ // ab-plane, with the calculated $\Delta S_M$ and $\Delta T_{ad}$ being comparable in both directions. Under $\Delta B = 2$ T, the extracted volumetric entropy change $-\Delta S_M=210$ mJ/K/cm$^3$ at ~ 1 K is considerably large compared to the $-\Delta S_M=126$ mJ/K/cm$^3$ in triangular-lattice frustrated $KBaGd(BO_3)_2$ and $-\Delta S_M=145$ mJ/K/cm$^3$ in benchmark magnetic refrigerant material $Gd_3Ga_5O_{12}$ (GGG) at ~1 K [18,22,40]. This large low-field MCE is appealing for its practical applications working in the low field region accessible by permanent magnet ($B ≤ 2$ T) rather than superconducting magnet. Under $\Delta B = 7$T, the maximum value of $-\Delta S_M=445$ mJ/K/cm$^3$ ($0.89R\ln 8$) exceeds most of $Gd^{3+}$ and $Eu^{2+}$ based ADR materials with effective S=7/2 moments as listed in Table 1 [18-25, 40-49].

Figure. 4(e) plots the expected $T_f$ as function of $T_i$. Starting from the initial condition ($T_i = 2$ K and $B_i = 7$T), $Gd_3BWO_9$ is expected to cool down to $T_f$ ~200 mK as field is driving back to 0 T. The attainable $T_f$~ 200 mK is lower than minimum temperature $T_{min} \sim 322$ mK for the hyperkagome magnet $Gd_3Ga_5O_{12}$ (GGG) magnet [40] and $T_{min} \sim 220$ mK for $NaGdP_2O_4$ [21]. As listed in Table 1, only the triangular-lattice frustrated $KBaGd(BO_3)_2$ realize a much lower $T_{min}$~ 122 mK than $Gd_3BWO_9$. But the achievable $T_{min}$~200 mK of $Gd_3BWO_9$ is well below the base temperature $T \sim 0.3$ K for the standard $^3$He refrigerators using self-contained and compact $^3$He cooler [50, 51]. Except for the achievable $T_{min}$ in the ADR process, $Gd_3BWO_9$ is also characterized by significant temperature change $\Delta T_{ad}$, defined by the difference between sample temperature ($T_s$) and based temperature ($T_0$) as denoted by the blue arrow in Fig.4(c). From the initial based condition ($T_0= 2$ K, $B_0= 0$ T), $\Delta T_{ad}$ reaches 20 K as field is increased up to field $B_s = 7$ T in an isentropic process. At different field intervals ($\Delta B=B_s-B_0$), the extracted $\Delta T_{ad}$ at different $\Delta B$ are plotted in Fig.4(f). Under $\Delta B = 7$ T, the derived $\Delta T_{ad}$ exceeds ~20 K in the temperatures ($T_0$) of 2-5 K, which is comparable with those of commercial GGG and $LiREF_4$(RE=Gd-Dy) coolants [31,32,41].

### 3.4 Test of magnetic cooling performance

The ADR cooling performance in sub-Kelvin temperatures was directly tested using a homemade sample stage inserted into the PPMS, the initial cooling conditions were set to $T_i = 2$

K and $B_i$ = 4 T or 6 T. From this starting point, magnetic field is swept down to the final field $B_f$ = 0 T in a rate of 0.3 T/min. During this procedure, the accessible sample temperature $T(B)$ is recorded under a quasi-adiabatic condition for $B$ // $c$-axis and $ab$-plane [see Fig. 5(a)]. From the temperature-field trajectories, the isentropic $T(B)$ curves at high field regimes ($B \geq 2$ T) are nearly overlapped along both field directions indicative of the isotropic field response in the FP state of $Gd_3BWO_9$, and it is confirmed by comparing the $T(B)$ data of polycrystals [ see Fig. S4]. At low fields ($B <$ 2 T), the $T(B)$ curves display anisotropic behavior with two distinct minimums for $B$ // $c$-axis and three broad valleys for $B$ // $ab$-plane [see the enlargement of $T(B)$ data in the inset of Fig. 5(a)]. After the demagnetization process from $B_i$= 6 T, the realized lowest $T_{min}$ ~168 mK ($B$ // $c$) and $T_{min}$ ~227 mK ($B$ // $ab$-plane) are lower than that of most Gd-based refrigerants as well as the recently reported Eu-based compounds [18-21, 43, 47, 48]. More interestingly, the detected temperature minimums in the adiabatic $T(B)$ curves appear at the finite fields instead of $B_f$ = 0 T. Checking the location of fields ($B_{c1}$ ~ 0.27 T and $B_{c2}$ ~ 0.96 T for $B$ // $c$, $B_{ab,1}$ ~ 0.092 T and $B_{ab,2}$ ~ 0.31 T for $B$ // $ab$-plane), we can find that they coincide with the ones extracted from the d$M$/d$B$ data. That means the minimal $T_{min}$ happen near the phase boundaries in the $B$-$T$ phase diagrams [ see Figs. 3(e, f)], signifying the presence of criticality-enhanced MCE as proposed in the low-dimensional quantum magnets [52, 53]. Compared to $B$ // $ab$-plane, the better cooling performance near $B_{c2}$ for $B$ // $c$-axis can be due to the presence of stronger spin fluctuations, which can lead to the much larger entropy release.To further quantify the MCE performance of $Gd_3BWO_9$ at sub-Kelvin temperatures, we have measured the holding time ($t_h$) of sample after ramping the field down to the final field $B_f$ = 0 T. When the external field starts to decrease from the initial condition ($T_i$ = 2 K, $B_i$ = 6 T), the time is defined to $t$ = 0. The evolution of sample temperature versus time $T(t)$ is recorded, the results for both single crystals and polycrystals are shown in Fig. 5(b). As seen, the holding time below ~250 mK is over an hour longer than the recently reports in Gd-based frustrated magnets $KBaGd(BO_3)_3$ and $NaGdP_2O_7$ [18,21]. The low attainable $T_{min}$ and long hold time of $Gd_3BWO_9$ let it as an excellent ADR material working in the sub-Kelvin temperatures.

Besides the good cooling performance in sub-Kelvin temperatures, it is instructive to characterize the MCE of $Gd_3BWO_9$ in extended temperature ranges for its application in continuous ADR system [23, 40, 54]. Thus, we carried out the adiabatic MCE measurements at given initial based temperature ($T_o$) under pulsed magnetic field up to ~32 T. Taking advantage of fast field-sweep rate, this method is a nearly adiabatic condition [31,55]. Fig. 5(c) shows the measured field dependence of sample temperatures ($T_s$) at various based temperatures, in which the calculated $T_s$ (open symbols) from the $S_M(T, B)$ data is also presented for comparison. The reasonable agreement at fields $B \leq 7$ T validate our MCE measurements under pulsed field. As denoted by the arrow in Fig. 5(c), $\Delta T_{ad}$ = 24-28 K is reached under $\Delta B$= 10 T in temperature intervals of $T_0$ = 2-10 K, which is close to $\Delta T_{ad}$ = 24-29 K of GGG in the comparable $T_0$ and $\Delta B$ [31], but $Gd_3BWO_9$ has more than 30% increase of magnetic entropy than that of GGG in this temperature region.

For RE-based ADR refrigerants working at sub-Kelvin temperatures, the Yb- and Gd-based magnets are relatively well studied. The former has the advantage on achieving lower $T_{min}$ down below 20 mK, the latter possesses larger cooling capacity but works at higher temperatures above 100 mK. Although pursuing the lowest possible $T_{min}$ is important, these two series of compounds are basically suitable for the different stages in a continuous ADR system [56]. The Yb-based materials are competitive in the first and second stages with temperature windows below 0.05 K and 0.045- 0.3 K, while the Gd-based magnets are more suitable in the third and fourth stages covering the temperature intervals of 0.25-1.1 K and 1.0-4.5 K. When the ADR cooling temperature regions is limited to 0.25-4.5 K, $Gd_3BWO_9$ exhibits excellent cooling performance among the representative Gd- and Eu-based refrigeration materials with $S_{eff}$=7/2 reported to date [18, 21, 40-50], as listed in Table 1. Considering the importance of cooling power per unit volume for practical application, the volumetric entropy density ($S_{GS}$/vol, calculated by dividing full entropy of magnetic ground state to unit cell volume) and the experimentally extracted -$\Delta S_m$ under field intervals ($\Delta B = B_i-B_f$) are used for the comparison. As seen from Table 1, $Gd_3BWO_9$ has larger $S_{GS}$/vol than the other Gd- and Eu-based compounds except $Gd_{9.33}[SiO_4]_6O_2$. Moreover, the low minimum temperature $T_{min}$ ~168 mK enables it to fully cover the temperature window of the third stage in continuous ADR system, in contrast to the other magnetically dense materials with $T_{min}$ remaining above 250 mK. Among the materials with achievable $T_{min}$ <250 mK, the values of $S_{GS}$/vol are much smaller than that of $Gd_3BWO_9$. The above comparative results highlight $Gd_3BWO_9$ as a superior magnetic refrigerant combining a quite large $S_{GS}$/vol and a relatively low $T_{min}$.

To further evaluate the ADR cooling performance, we compare the extracted $\Delta S_M$ in extended temperature ranges (0.5-4 K) of $Gd_3BWO_9$ with other representative materials as listed in Table 2. The $\Delta S_M$ of $Gd_3BWO_9$ under $\Delta B$ = 7 T are comparable with the values of $Gd_{9.33}[SiO_4]_6O_2$ and far exceeds those of other materials in temperature ranges of 2-4 K. Below 1 K, however, the $\Delta S_M$ values are lower than those of $LiGdF_4$ and $Gd_{9.33}[SiO_4]_6O_2$ partially due to the reduced ordering temperatures $T_o$~0.48 K in $LiGdF_4$ [41] and the spin-disordered ground state in $Gd_{9.33}[SiO_4]_6O_2$ with the short-range spin correlation established at $T_{sr}$~0.52 K [25]. In this compound, the $Gd^{3+}$ ions are randomly occupied on the frustrated one-dimensional spin chain with 2/3 occupation ratio, thus both geometrical spin frustration and structural disorder effect are responsible for the suppression of long-range magnetic order and the correspondingly enhanced sub-Kelvin MCE. However, the presence of structural disorder complicates the understanding of how geometric frustration alone regulates the MCE behavior. In this respect, $Gd_3BWO_9$ stands out as a disorder-free frustrated system. The large differences in ionic radius of magnetic $Gd^{3+}$ and nonmagnetic $B^{3+}/W^{6+}$ cations prevent the antisite chemical disorder [26,27] in this material, making it a clean system for examining the role of geometric frustration on optimizing the sub-Kelvin MCE.

Besides the large $S_{GS}$/vol, Gd$_3$BWO$_9$ exhibits a significantly low $T_{min}$~168 mK compared to its $T_o$~ 1.0 K. To highlight the large difference between $T_{min}$ and $T_o$ of Gd$_3$BWO$_9$, we summarize the comparative results on the reported sub-Kelvin Gd-and Eu- based ADR materials[18,22-25, 41,43,45,47], and plot the $T_o$ and $T_{min}$ versus $N_{mag}$ of materials in Fig. 6. Here, $T_o$ denotes the long-range AFM ordering ($T_N$) or the onset of short-range magnetic order ($T_{sr}$), and $T_{min}$ is the lowest attainable temperature after ADR process from initial condition of $T_i$ = 2 K and $B_i$ = 3- 12 T [see the details in Table 1]. A general trend is that larger $N_{mag}$ induces higher $T_o$ with a roughly linear dependence, which is reasonable due to the increased magnetic exchange and dipolar interactions. On the other side, $T_{min}$ increase relatively slower with the variation of $N_{mag}$. Thus, identifying materials with large $N_{mag}$ and low $T_o$ is challenging but essential for improving the sub-Kelvin ADR cooling performance. Considering that the benchmark refrigerant GGG has $N_{mag}^{GGG}$=12.65 nm$^{-3}$ and $T_o^{GGG}$=0.8 K [56], the identification of the materials with $N_{mag} > N_{mag}^{GGG}$ and $T_0 < T_o^{GGG}$ offers a promising route to surpass commercial GGG as the next-generation sub-Kelvin ADR coolants. To satisfy the above criteria, the use of geometrically frustrated magnets (GFMs) is effective to simultaneously increase $N_{mag}$ and lower $T_o$ because the $T_o$ can be reduced or even suppressed by spin frustration compared to the unfrustrated materials. For Gd$_3$BWO$_9$, its $T_o$ = 1 K is already reduced by spin frustration compared to its exchange energy scale ~3 K [see the estimation based on magnetization data in Fig. S2], though being still higher than $T_o^{GGG}$. Thus, the realization of lower $T_{min}$ ~168 mK than $T_{min}^{GGG}$~322 mK in the same ADR cooling condition is quite intriguing.

A detailed analysis of magnetic entropy $S_M(T,B)$ landscape is crucial for understanding the mechanism behind the exceptionally low $T_{min}$ in Gd$_3$BWO$_9$. In the ADR runs, the coolants experience a quasi-isentropic process from the initial condition ($T_i$, $B_i$) to the final condition ($T_f$, $B_f$). Once the $B_i$ and $B_f$ are fixed, $\Delta T_{ad}$ = $T_i$-$T_f$ is fully determined by the entropy profiles of $S_M(T, B_i)$ and $S_M(T, B_f)$ in the $S_M(T, B)$ curves [see Fig. 4(c)]. Since a larger $\Delta T_{ad}$= $T_i$-$T_f$ corresponds to a lower $T_{min}$, two approaches can be employed to reduce $T_{min}$. One is to realize a smaller value of entropy $S_M(T_i = 2K, B = B_i)$ at the starting point, the other is to retain a large entropy at the lower temperatures, which will shift the $T_f$ of endpoint to lower temperature in the $S_M(T,B)$ phase diagram. For the former approach, the weak mutual interactions among Gd$^{3+}$ ions in Gd$_3$BWO$_9$ lead to the fast shift of $C_M(T)$ curves in field to high temperatures and correspondingly small value of $S_M(T_i, B_i)$. This is because that the movement of Schottky-like $C_M(T)$ feature roughly follows $gS_{eff}\mu_B(B-B_s)$ [17,18]. At the initial cooling condition of $T_i$ = 2K and $B_i$ = 7T, the integrated $S_M(T_i, B_i)$~0.035$R$ln8 of Gd$_3$BWO$_9$ is smaller than ~0.052$R$ln8 of Gd$_{9.33}$[SiO$_4$]$_6$O$_2$ [25] and 0.069$R$ln8 of NaGdP$_2$O$_7$[21]. For the latter approach, by setting $B_f$ = 0 T, since the attainable $T_{min}$ of Gd$_3$BWO$_9$ is well below $T_N$, the value of $T_{min}$ is mainly determined by the zero-field $C_M(T)$ data. From this viewpoint, the slow drop of $C_M \propto T^2$ ($T$ < 0.37 K) rather than $C_M \propto T^3$ is favourable for the accumulation of $S_M$ below $T_f$ [ see Fig. 2(a)]. Thus, not only the lower $T_o$ but also the slower decay of $C_M/T$ below $T_o$ can facilitate a large accumulation of $S_M(T_f, B_f)$, leading to a lower $T_{min}$ under $B_f$

= 0 T.

Tuning materials toward a quantum critical point can generate an additional enhancement of spin fluctuations, leading to a marked temperature decrease under the adiabatic condition. Checking the $T$-dependence of $C_M/T$ at different fields [see the inset of Fig. 4(a)], we can find that the values of $C_M/T$ at 1 T are larger than those measured at fields below or above 1T in low temperature regions ($T < T^*$). This means the integrated entropy $S_M(T, B=B_f)$ reaches maxima if setting $B_f$ = 1 T. This can explain the lowest $T_{min}$~168 mK ($B // c$-axis) is realized near the critical field $B_{c2}$ ~0.96 T in the ADR cooling performance test [see Fig. 5(a)]. Thus, the lowest $T_{min}$~168 mK realized at $B_f = B_{c2}$ rather than $T_{min}$~232 mK at $B_f$ = 0 T is facilitated by the criticality-enhanced MCE. Compared to the zero-field spin fluctuation originating from geometric spin frustration of Kagome lattice, the increased spin fluctuations near quantum critical point induce an additional accumulation of entropy responsible for the ~26% decrease from $T_{min}$~227 mK at $B_f$ = 0 T. This mechanism, known as criticality-enhanced MCE, has been proposed for low dimensional frustrated magnets, especially the novel spin disordered system as a mean to improve the cooling performance, as realized in the spin supersolid candidate $Na_2BaCo(PO_4)_2$ with a quite low $T_{min}$ ~94 mK that is attained near its quantum critical point (QCP)[13]. The present cooling performance results of $Gd_3BWO_9$ reveal that the criticality-enhanced MCE can also be used to optimize the cooling performance of Gd-based frustrated antiferromagnets.

## 4. Conclusions

In summary, we have reported the anisotropic magnetic phase diagram and magnetic cooling performance at sub-Kelvin temperatures of Kagome magnet $Gd_3BWO_9$. This material demonstrates an excellent MCE performance in sub-Kelvin regime, featured by large volumetric magnetic entropy in a wide temperature window from ~0.25-4 K, remarkable low $T_{min}$~168 mK and long holding time than an hour following the ADR process. Compared to most of dense Gd-based antiferromagnets, the attainable quite low $T_{min}$~168 mK in $Gd_3BWO_9$ is beneficial from the enhancement of spin fluctuations arising from the geometric spin frustration and field-tuned criticality effect, which together optimize the MCE at sub-Kelvin temperatures. Our work thus provides a realistic example of utilizing field-tunable criticality to improve the ADR cooling performance in frustrated magnets.

**Acknowledgements**

The work is supported by the National Key Research and Development Program (Grant No. 2024YFA1611200 & 2023YFA1406500). This work was supported by the National Natural Science Foundation of China (Grant No. 12141002 and 52088101) and the Strategic Priority Research Program of the Chinese Academy of Sciences (XDB1270000). A portion of this work was carried out at the synergetic extreme condition user facility (SECUF). We would like to thank the staff of the analysis center of Huazhong University of Science and Technology for their assistance in structural characterizations.

**Author Declarations**

The authors have no conflicts to disclose.

Table 1. Comparison of key parameters for Gd- and Eu-based ADR materials at subKelvin temperatures: To is the magnetic ordering temperature and short range order is labeled by (*), Tmin is the attainable minimum temperature after the ADR process from the initial condition of ($T_i$, $B_i$), $N_{mag}$ is the density of magnetic ions, $S_{GS}$ is the entropy of the ground-state multiplet, and $S_{GS}$/vol is the volumetric magnetic entropy density, and $-\Delta S_M$ is the magnetic entropy change in 1 K and $\Delta B$ = 6 T or 7 T.

| Materials | $T_o$ (K) | $T_{min}$ (mK) | $T_i, B_i$ (K, T) | $N_{mag}$ (nm$^{-3}$) | $S_{GS}$ | $S_{GS}$/vol. [mJ/(K cm$^3$)] | $-\Delta S_M$ [mJ/(K cm$^3$)] | References |
|---|---|---|---|---|---|---|---|---|
| **Gd$_3$BWO$_9$** | 1 | 168 | 2,6 | **17.49** | $R$ln8 | 502.2 | 303 | **This work** |
| Gd$_3$Ga$_5$O$_{12}$ | 0.8 (*) | 322 | 2,6 | 12.65 | $R$ln8 | 363.2 | 274 | [22] |
| KBaGd(BO$_3$)$_2$ | 0.263 | 122 | 2,5 | 6.46 | $R$ln8 | 185.8 | 149 | [18] |
| Gd$_2$Ti$_2$O$_7$ | 1.0 | 480 | 2,12 | 15.14 | $R$ln8 | 435 | 115 | [24] |
| K$_2$GdNb$_5$O$_{15}$ | 0.31 | 175 | 2,5 | 3.29 | $R$ln8 | 93.2 | 74 | [44] |
| GdPO$_4$ | 0.77 | 325 | 2,7 | 14.47 | $R$ln8 | 415.6 | 290 | [43] |
| Gd(HCOO)$_3$ | 0.8 | 356 | 2,3 | 7.94 | $R$ln8 | 228.1 | 192 | [42] |
| Gd$_{9.33}$[SiO4]$_6$O$_2$ | 0.5(*) | 210 | 2,7 | 17.74 | $R$ln8 | 509.4 | 344 | [25] |
| NaGdP$_2$O$_7$ | 0.57 | 220 | 2,5 | 7.3 | $R$ln8 | 210 | 93.2 | [21] |
| LiGdF$_4$ | 0.48 | <500 | 2,5 | 13.36 | $R$ln8 | 383.8 | 316 | [41] |
| EuCl$_2$ | 1.3 | 346 | 2,5 | 13.15 | $R$ln8 | 377.1 | 151 | [47] |
| EuB$_4$O$_7$ | 068 | 289 | 2,4 | 9.911 | $R$ln8 | 284.5 | 186 | [48] |

Table 2. Comparison of $T_o$, $T_{min}$, $S_{GS}$/vol and $-\Delta S_M$ under different $\Delta B=B_f-B_0$ for the representative Gd- and Eu-based ADR materials in selected temperature ranges from 0.5 K to 4 K.

| Materials | $T_o$ (K) | $T_{min}$ (mK) | $S_{GS}$/vol. [mJ/(K cm$^3$)] | $-\Delta S_M$ [mJ/(K cm$^3$)] | | | | References |
|---|---|---|---|---|---|---|---|---|
| | | | | $\Delta B$ (T) | 0.5 K | 1 K | 2 K | 4 K | |
| **Gd$_3$BWO$_9$** | 1 | 170 | 502.2 | 7 | 110 | 303 | 438 | 408 | **This work** |
| $Gd_{9.33}[SiO_4]_6O_2$ | 0.5 | 300 | 509.4 | 7 | 183 | 344 | 440 | 391 | [25] |
| $LiGdF_4$ | 0.48 | <500 | 383.8 | 6 | - | 316 | 348 | 290 | [41] |
| $Gd_3Ga_5O_{12}$ | 0.8(*) | 322 | 363.2 | 6 | 108 | 274 | 318 | 260 | [22] |
| $Gd(OH)CO_3$ | 0.7 | 508 | 394.7 | 7 | 106 | 280 | 348 | 317 | [45] |
| $Gd_2Ti_2O_7$ | 1.0 | 480 | 435 | 7 | 24.9 | 115 | 190 | 184 | [24] |
| $GdPO_4$ | 0.77 | 385 | 415.6 | 7 | 62 | 290 | 375 | 332 | [43] |
| $KBaGd(BO_3)_2$ | 0.263 | 122 | 185.8 | 5 | 118 | 149 | 143 | 130 | [18] |
| $EuCl_2$ | 1.3 | 346 | 377.1 | 5 | 74 | 151 | 369 | 270 | [47] |

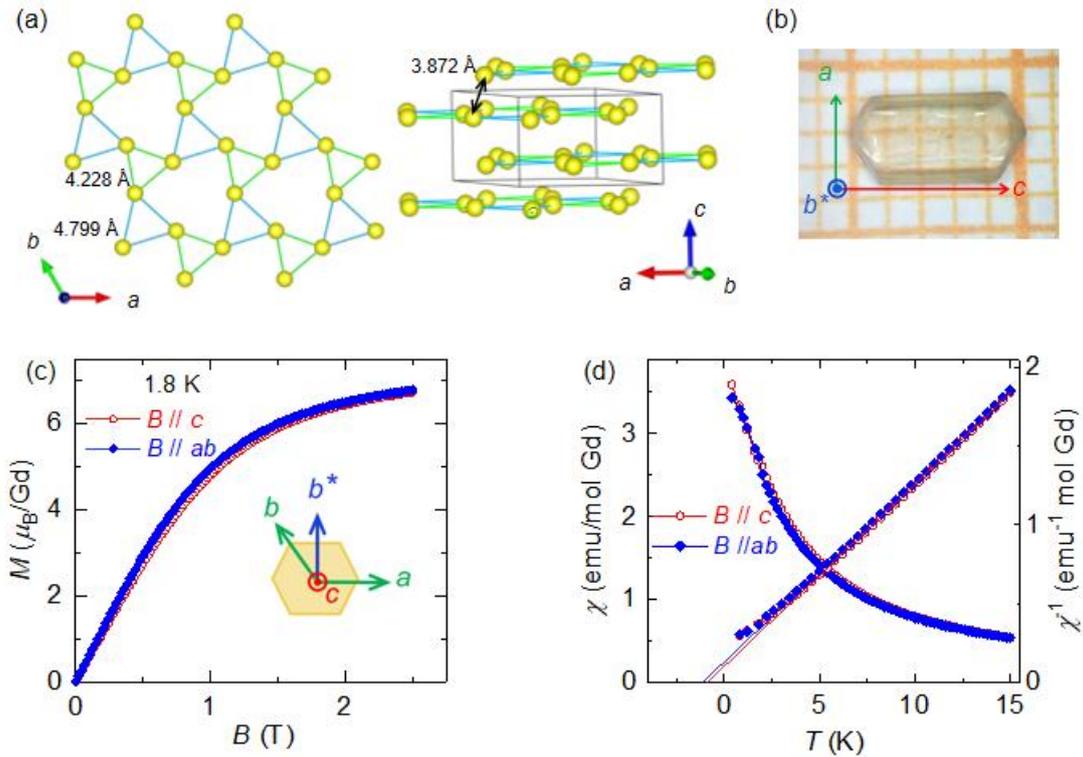

Fig. 1(a) The Kagome lattice formed by $Gd^{3+}$ ions within the *ab*-plane and its stacking along the *c*-axis in $Gd_3BWO_9$. (b) Picture of a typical $Gd_3BWO_9$ single crystal. (c) Isothermal magnetization *M(B)* curves of $Gd_3BWO_9$ at 1.8 K with *B // c* axis and *B // ab*-plane, respectively. Inset shows the definition of axes. (d) Temperature dependence of magnetic susceptibility $\chi$ (T) and inverse magnetic susceptibility $1/\chi$(T) under *B* = 0.1 T with *B // c* axis and *B // ab*-plane, respectively.

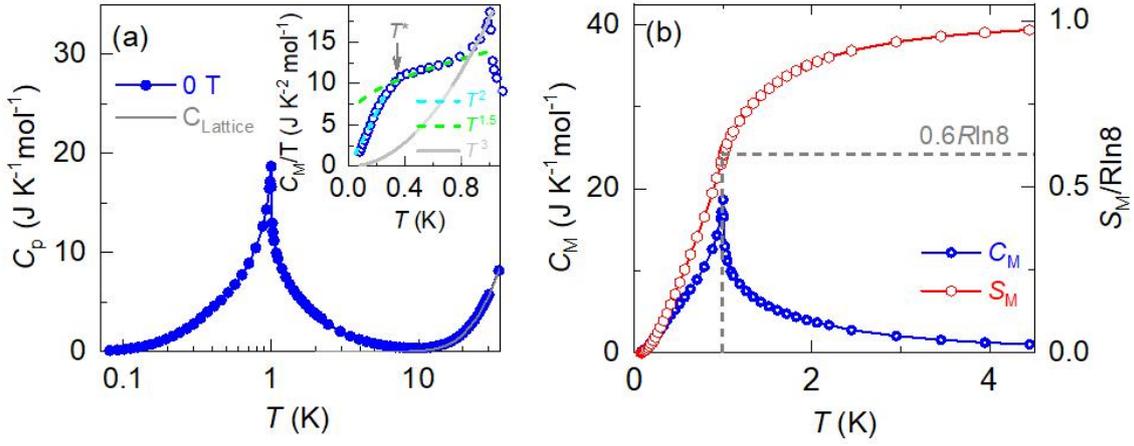

Fig. 2 (a) Zero field specific heat $C_p(T)$ curves of Gd$_3$BWO$_9$ and its isostructural analog Eu$_3$BWO$_9$ for lattice contribution to $C_p(T)$, inset shows the power law fit on the $C_M/T$ versus $T$. (b) Magnetic specific heat $C_M(T)$ and entropy $S_M(T)$ data of Gd$_3$BWO$_9$.

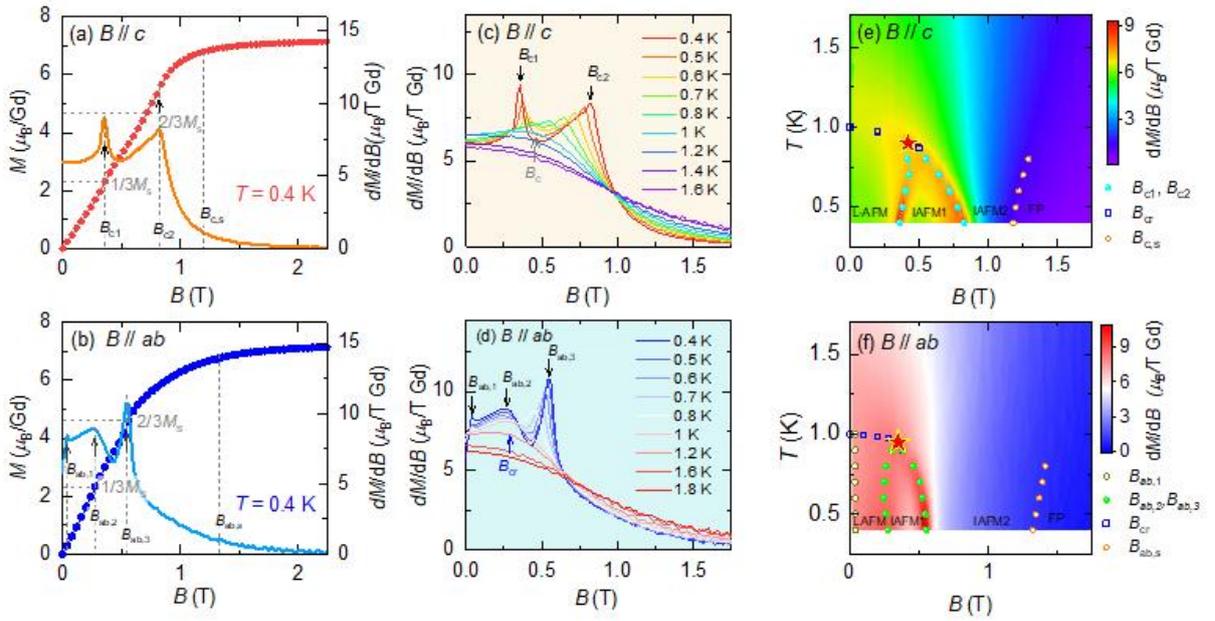

Fig. 3 (a, b) The $M(B)$ and $dM/dB$ curves of Gd$_3$BWO$_9$ at 0.4 K with $B \parallel c$ axis and $B \parallel ab$-plane, respectively. (c, d) The $dM/dB$ curves at different temperatures (0.4 K ≤ $T$ ≤ 1.8 K) along $c$-axis and $b^*$-axis. (e, f) The $B$-$T$ magnetic phase diagrams constructed by the $dM/dB$ data for $B$ along $c$ axis and $b^*$ axis, the stars denote the multi-critical points in the phase diagram.
141414

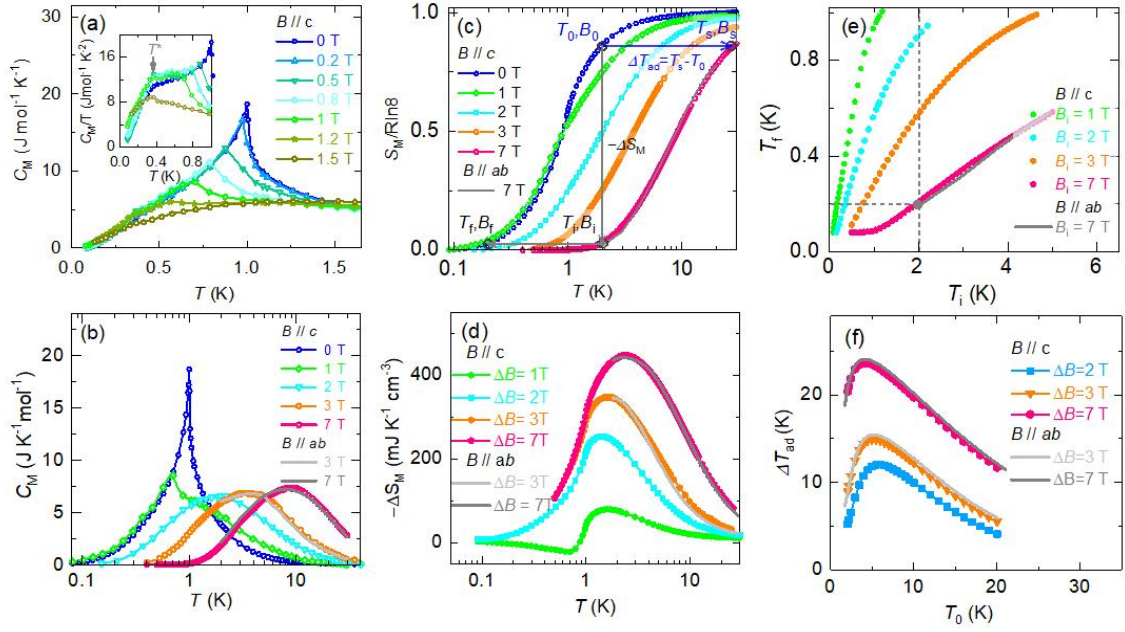

Fig. 4 (a) Magnetic specific heat $C_M(T)$ curves of $Gd_3BWO_9$ at low magnetic fields ($B \leq 1.5$ T) for $B \parallel c$, inset shows the $C_M/T$ data plotted versus $T$. (b) Magnetic specific heat $C_M(T)$ curves of $Gd_3BWO_9$ at selected magnetic fields. (c) The resultant magnetic entropy $S_M(B,T)$ under different magnetic fields of $Gd_3BWO_9$, the arrows denote the isothermal suppression of entropy $-\Delta S_M$ and adiabatic temperature change $\Delta T_{ad}$ in the ADR processes. (d) Temperature dependence of $-\Delta S_M(T)$ at different field changes ($\Delta B$). (e) The final temperature $T_f$ versus initial temperature $T_i$ cooled from different initial field calculated from the $S_M(B,T)$ results. (f) The $T_t$ versus $T_0$ at different $\Delta B$ in an ADR processes.

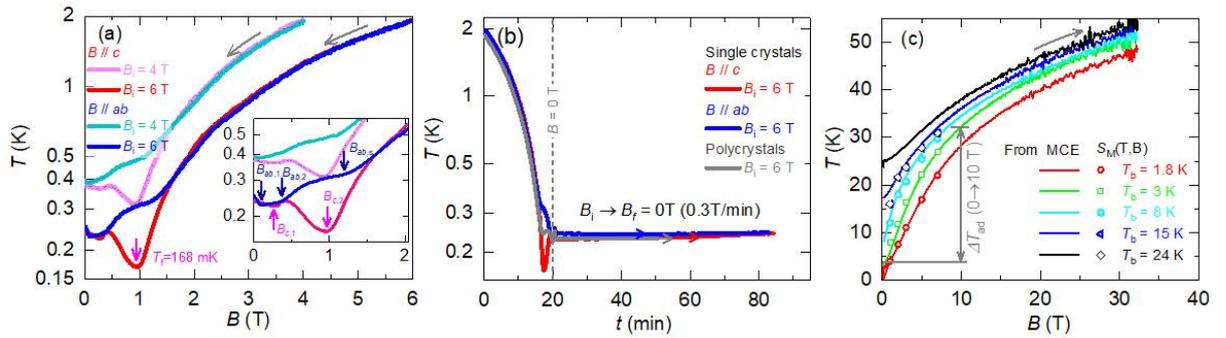

Fig. 5 (a) The quasi-adiabatic demagnetization refrigeration test of $Gd_3BWO_9$ single crystals (0.7 g) started from two different initial cooling conditions ($T_i = 2$ K, $B_i = 4$ T) and ($T_i = 2$ K, $B_i = 6$ T) for $B \parallel c$ axis and $B \parallel ab$-plane, (b) Comparison of the holding time of $Gd_3BWO_9$ single crystals and polycrystals under same initial condition $T_i = 2$ K, $B_i = 6$ T, (c) Field dependence of temperature change $T_t$ versus $T_0$ of $Gd_3BWO_9$ single crystals measured with pulsed magnetic fields up to 32 T applied along $c$-axis, the open symbols show the deduced $T_t$ from the specific heat data, the arrow corresponds to $\Delta T_{ad}$ (0→ 10 T).

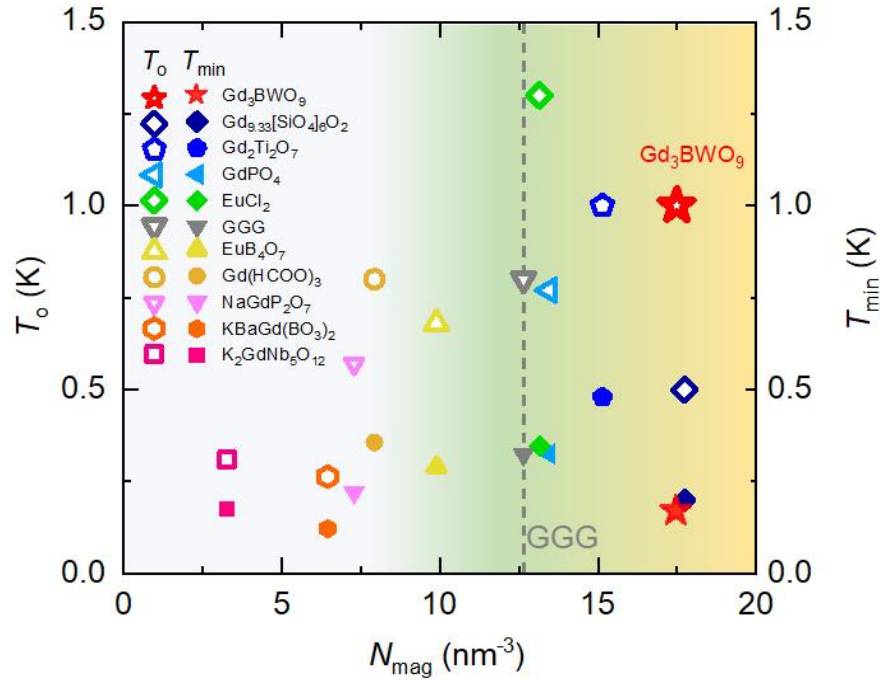

Fig. 6 Comparison of magnetic ordering temperature ($T_o$) and minimum cooling temperature ($T_{min}$) versus magnetic ion density ($N_{mag}$) on the representative Gd and Eu-based magnetocaloric materials.